\documentclass[twocolumn,preprintnumbers,amsmath,amssymb]{revtex4-1}

\usepackage{graphicx}
\usepackage{dcolumn}
\usepackage{bm}
\usepackage{rotating}

\usepackage{bbm}
\usepackage{verbatim}
\usepackage{afterpage}

\usepackage{color}
\definecolor{lightblue}{rgb}{0.2,0.2,0.7}
\definecolor{darkblue}{rgb}{0,0.25,0.5}
\definecolor{redbrown}{rgb}{0.875,0.25,0.125}
\definecolor{darkgreen}{rgb}{0,0.5,0}

\newcommand{\bra}[1]{\ensuremath{\langle #1 \vert}}
\newcommand{\ket}[1]{\ensuremath{\vert #1  \rangle}}

\renewcommand{\b}[1]{\ensuremath{\mathbf{#1}}}
\renewcommand{\H}{\ensuremath{\text{H}}}
\renewcommand{\l}{\ensuremath{\lambda}}

\newcommand{\LDA}{\ensuremath{\text{LDA}}}
\newcommand{\GGA}{\ensuremath{\text{GGA}}}
\newcommand{\HF}{\ensuremath{\text{HF}}}

\newcommand{\n}{\ensuremath{\nabla}}

\begin{document}

\title{Double-hybrid density-functional theory made rigorous}

\author{Kamal Sharkas$^{1,2}$}\email{kamal.sharkas@etu.upmc.fr}
\author{Julien Toulouse$^1$}\email{julien.toulouse@upmc.fr}
\author{Andreas Savin$^1$}\email{savin@lct.jussieu.fr}
\affiliation{
$^1$Laboratoire de Chimie Th\'eorique, Universit\'e Pierre et Marie Curie and CNRS, 75005 Paris, France\\
$^2$Atomic Energy Commission of Syria, P.O. Box 6091, Damascus, Syria}


\date{\today}

\begin{abstract}
We provide a rigorous derivation of a class of double-hybrid approximations, combining Hartree-Fock exchange and second-order M{\o}ller-Plesset correlation with a semilocal exchange-correlation density functional. These double-hybrid approximations contain only one empirical parameter and use a density-scaled correlation energy functional. Neglecting density scaling leads to a one-parameter version of the standard double-hybrid approximations. We assess the performance of these double-hybrid schemes on representative test sets of atomization energies and reaction barrier heights, and we compare to other hybrid approximations, including range-separated hybrids. Our best one-parameter double-hybrid  approximation, called 1DH-BLYP, roughly reproduces the two parameters of the standard B2-PLYP or B2GP-PLYP double-hybrid approximations, which shows that these methods are not only empirically close to an optimum for general chemical applications but are also theoretically supported.
\end{abstract}

\maketitle

\section{Introduction}

Density-functional theory (DFT)~\cite{HohKoh-PR-64,KohSha-PR-65,Koh-RMP-99} is a powerful approach for electronic-structure calculations of atoms, molecules and solids. In its Kohn-Sham (KS) formulation, a series of approximations for the exchange-correlation energy have been developed for an ever-increasing accuracy: local density approximation (LDA), semilocal approximations (generalized-gradient approximations (GGA) and meta-GGA), hybrid approximations introducing Hartree-Fock (HF) exchange, and nonlocal correlation approximations using virtual KS orbitals~\cite{KumKro-RMP-08}.

In this context, Grimme~\cite{Gri-JCP-06} recently introduced the family of so-called {\it double-hybrid} (DH) density-functional approximations which mix HF exchange with a semilocal exchange density functional and second-order M{\o}ller-Plesset (MP2) correlation with a semilocal correlation density functional:
\begin{eqnarray}
E_{xc}^{\text{DH}} &=& a_x E_{x}^{\HF} + (1-a_x) E_{x}[n]
\nonumber\\
&& + (1-a_c) E_{c}[n] + a_c E_{c}^{\text{MP2}},
\label{DH}
\end{eqnarray}
where the first three terms are calculated in an usual self-consistent hybrid KS calculation, and the last perturbative term evaluated with the previously obtained orbitals is added \textit{a posteriori}. The B2-PLYP double-hybrid approximation~\cite{Gri-JCP-06} is obtained by choosing the Becke 88 (B) exchange functional~\cite{Bec-PRA-88} for $E_{x}[n]$ and the Lee-Yang-Parr (LYP) correlation functional~\cite{LeeYanPar-PRB-88} for $E_{c}[n]$, and the empirical parameters $a_x=0.53$ and $a_c=0.27$ optimized for the G2/97 subset of heats of formation. The mPW2-PLYP double-hybrid approximation~\cite{SchGri-PCCP-06} uses the modified Perdew-Wang (mPW) exchange functional~\cite{AdaBar-JCP-98}, and has very similar optimized parameters $a_x=0.55$ and $a_c=0.25$. These two double-hybrid approximations reach on average near-chemical accuracy for the thermodynamical data of the G3/05 set~\cite{SchGri-PCCP-06}. Similar double-hybrid approximations have also been obtained by reoptimizing the parameters $a_x$ and $a_c$ for a spin-restricted open-shell version of the method~\cite{GraMenGoeGriRad-JPCA-09} or for different test sets~\cite{TarKarSerVuzMar-JPCA-08,KarTarLamSchMar-JPCA-08}. In particular, targeting both thermochemistry and kinetics applications has given the reoptimized parameters $a_x=0.65$ and $a_c=0.36$ which defines the general-purpose B2GP-PLYP double-hybrid approximation~\cite{KarTarLamSchMar-JPCA-08}. The so-called multicoefficient correlation methods combining HF, DFT and MP2 energies can also be considered to be a form of double-hybrid approximation~\cite{ZhaLynTru-JPCA-04,ZhaLynTru-PCCP-05,ZheZhaTru-JCTC-09}, and the connection was made explicit in Ref.~\onlinecite{SanPer-JCP-09}. Three- or four-parameter double-hybrid approximations have also been proposed~\cite{ZhaXuGod-PNAS-09,ZhaLuoXu-JCP-10,ZhaLuoXu-JCP-10b}, scaling differently the LDA and GGA components of the density functionals, in the style of the first hybrid DFT approximations~\cite{Bec-JCP-93}. For systems with van der Waals interactions, good accuracy can be obtained by further adding an empirical dispersion term~\cite{SchGri-PCCP-07} or by increasing the amount of MP2 correlation at long interelectronic distances~\cite{BenDisLocChaHea-JPCA-08}. 

Although the above-mentioned double-hybrid approximations yield very promising results and are already largely used, they suffer from a lack of theoretical justification. It has been tried~\cite{ZhaXuGod-PNAS-09} to motivate these approaches by invoking the adiabatic connection formalism~\cite{Har-PRA-84} and second-order G\"orling-Levy perturbation theory (GL2)~\cite{GorLev-PRB-93}, but several unjustified empirical steps remain (e.g., dropping the single-excitation term in the GL2 expression). On the contrary, the range-separated double-hybrid RSH+lrMP2 method of Ref.~\onlinecite{AngGerSavTou-PRA-05}, which combines long-range HF exchange and long-range MP2 correlation with a short-range exchange-correlation density functional, has been rigorously derived using the formally exact multideterminant extension of the Kohn-Sham scheme based on range separation. In this work, we apply an analogous formalism without range separation which leads to a rigorous derivation of a form of double-hybrid approximation. In this double-hybrid scheme, only one empirical parameter appears, the appropriate correlation energy functional is obtained by uniform coordinate scaling of the density, and the MP2 correlation energy expression appears naturally without the need to neglect single-excitation contributions. We test the proposed double-hybrid scheme on representative sets of atomization energies and reaction barrier heights, and compare with other hybrid approximations.

\section{Theory}

We consider the usual adiabatic connection of DFT linking the non-interacting Kohn-Sham Hamiltonian ($\l=0$) to the exact Hamiltonian ($\l=1$) by linearly switching on the Coulombic electron-electron interaction $\l \hat{W}_{ee}$~\cite{Har-PRA-84},
\begin{eqnarray}
\hat{H}^\l = \hat{T}+\hat{V}_{\text{ext}}+\l\hat{W}_{ee} + \hat{V}_{\H xc}^{\l}[n],
\label{Hl}
\end{eqnarray} 
where $\hat{T}$ is the kinetic energy operator, $\hat{V}_{\text{ext}}$ is a scalar external potential operator (e.g., nuclei-electron), and $\hat{V}_{\H xc}^{\l}[n]$ is the Hartree-exchange-correlation potential operator keeping the one-electron density $n$ constant for all values of the coupling constant $\l \geq 0$. Using the formalism of the multideterminant extension of the Kohn-Sham scheme (see, e.g., Refs.~\onlinecite{TouColSav-PRA-04,AngGerSavTou-PRA-05}), for any $\l$, the {\it exact} energy can be expressed as the following minimization over multideterminant wave functions $\Psi$:
\begin{eqnarray}
E &=& \min_{\Psi} \Bigl\{\bra{\Psi}\hat{T}+\hat{V}_{\text{ext}}+\l\hat{W}_{ee} \ket{\Psi}+\bar{E}_{\H xc}^{\l}[n_{\Psi}]\Bigl\}, 
\label{EminPsi}
\end{eqnarray} 
where $n_{\Psi}$ is the density coming from $\Psi$ and $\bar{E}_{\H xc}^{\l}[n]=E_{\H xc}[n]-E_{\H xc}^{\l}[n]$ is the {\it complement} $\l$-dependent Hartree-exchange-correlation density functional, i.e. the difference between the usual Kohn-Sham density functional $E_{\H xc}[n]$ and the $\l$-dependent density functional $E_{\H xc}^{\l}[n]$ corresponding to the Hamiltonian~(\ref{Hl}). This complement density functional generates the potential in Eq.~(\ref{Hl}) which keeps the density constant: $\hat{V}_{\H xc}^{\l}[n]=\int d\b{r} \, \hat{n}(\b{r}) \, \delta \bar{E}_{\H xc}^{\l}[n]/\delta n(\b{r})$, where $\hat{n}(\b{r})$ is the density operator. Since the Hartree and exchange contributions are first order in the electron-electron interaction, their dependence on $\l$ is just linear,
\begin{eqnarray}
\bar{E}_{\H}^{\l}[n]=(1-\l) E_{\H}[n],
\label{}
\end{eqnarray} 
\begin{eqnarray}
\bar{E}_{x}^{\l}[n]=(1-\l) E_{x}[n],
\label{}
\end{eqnarray} 
where $E_{\H}[n]$ and $E_{x}[n]$ are the usual Kohn-Sham Hartree and exchange density functionals. The correlation contribution is not linear in $\l$ but can be obtained by uniform coordinate scaling of the density~\cite{LevPer-PRA-85,LevYanPar-JCP-85,Lev-PRA-91,LevPer-PRB-93},
\begin{eqnarray}
\bar{E}_{c}^{\l}[n] &=& E_{c}[n]-E_{c}^{\l}[n]
\nonumber\\
                    &=& E_{c}[n]-\l^2 E_{c}[n_{1/\l}],
\label{}
\end{eqnarray} 
where $E_{c}[n]$ is the usual Kohn-Sham correlation functional, $E_{c}^{\l}[n]$ is the correlation functional corresponding to the Hamiltonian~(\ref{Hl}), and $n_{1/\l}(\b{r})=(1/\l)^3 n(\b{r}/\l)$ is the scaled density.

To avoid possible confusions with previous work, we note that numerous exchange-correlation functional approximations have been constructed using the adiabatic-connection integral formula (see, e.g., Refs.~\onlinecite{Bec-JCP-93a,LevMarHan-JCP-96,Ern-CPL-96,GriLeeBae-IJQC-96,BurErnPer-CPL-97,ErnPerBur-IJQC-97,SeiPerLev-PRA-99,SeiPerKur-PRA-00,SeiPerKur-PRL-00,MorCohYan-JCP-06a,CohMorYan-JCP-07,CohMorYan-JCP-07b}), $E_{xc}[n] = \int_0^1 U_{xc,\alpha} \, d\alpha$ where $U_{xc,\alpha}$ is the integrand (exchange + potential correlation energy) that needs to be approximated. In the same language, the complement exchange-correlation functional used in the present work would write $\bar{E}_{xc}^{\l}[n] = \int^1_\lambda U_{xc,\alpha} \, d\alpha$. The confusion is possible because $U_{xc,\alpha}$ has sometimes been called $E_{xc,\alpha}$ in the literature. Although we do not use in practice this adiabatic-connection integral, it should help to clarify that in this work the coupling constant $\l$ is fixed and the complement correlation functional $\bar{E}_{c}^{\l}[n]$ does include a kinetic correlation energy contribution.

\begin{figure*}
\includegraphics[scale=0.30,angle=-90]{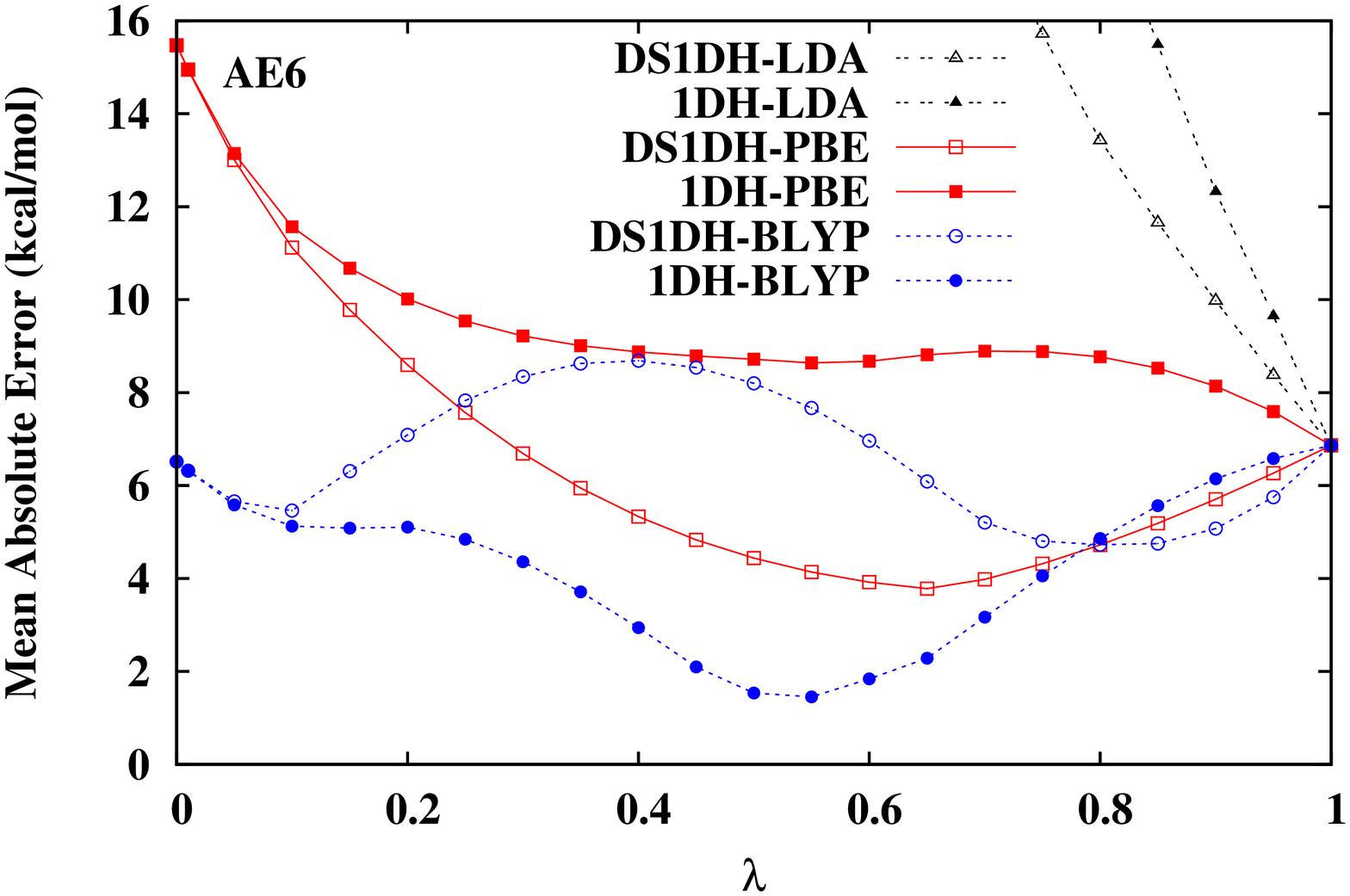}
\includegraphics[scale=0.30,angle=-90]{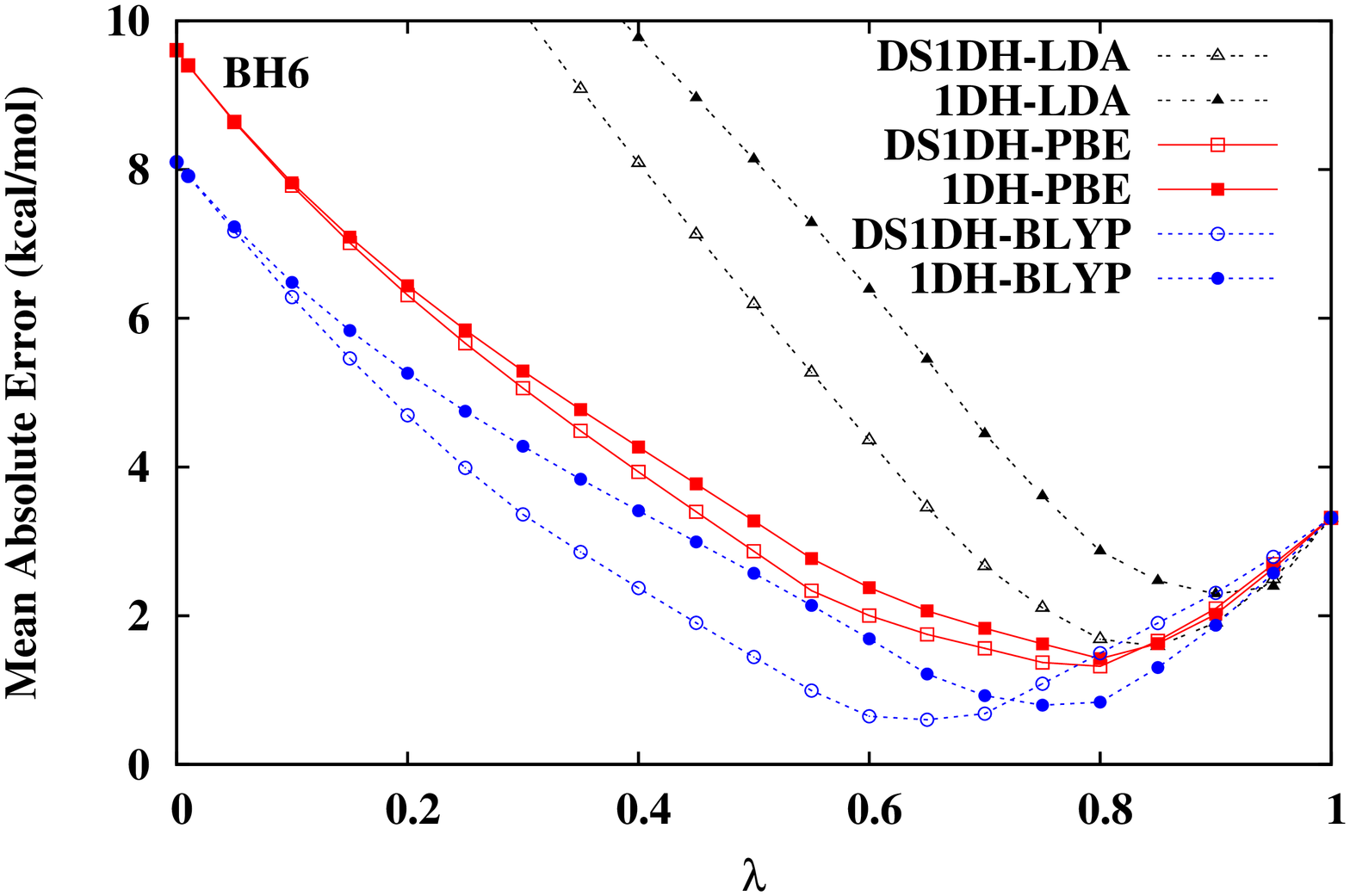}
\caption{(Color online) MAEs for the AE6 (left) and BH6 (right) test sets as functions of the parameter $\l$ for the 1DH and DS1DH approximations with LDA, PBE, and BLYP exchange-correlation density functionals. All calculations were carried out with the cc-pVQZ basis set.
}
\label{fig:AE6BH6}
\end{figure*}

We now define a {\it density-scaled one-parameter hybrid} (DS1H) approximation by restricting the minimization in Eq.~(\ref{EminPsi}) to single-determinant wave functions $\Phi$:
\begin{eqnarray}
E^{\text{DS1H},\l} &=& \min_{\Phi} \Bigl\{\bra{\Phi}\hat{T}+\hat{V}_{\text{ext}}+\l\hat{W}_{ee} \ket{\Phi}+\bar{E}_{\H xc}^{\l}[n_{\Phi}]\Bigl\}, 
\nonumber\\
\end{eqnarray} 
obtaining an energy which necessarily depends on $\l$. The minimizing single-determinant wave function $\Phi^\l$ is calculated by the self-consistent eigenvalue equation:
\begin{eqnarray}
\left( \hat{T}+\hat{V}_{\text{ext}}+\l \hat{V}_{\H x}^{\HF}[\Phi^\l] + \hat{V}_{\H xc}^{\l}[n_{\Phi^\l}] \right) \ket{\Phi^\l} = {\cal E}_0^{\l} \ket{\Phi^\l},
\label{DS1Heigenval}
\end{eqnarray} 
where $\hat{V}_{\H x}^{\HF}[\Phi^\l]$ is the nonlocal HF potential operator evaluated with the DS1H wave function $\Phi^\l$ and $\hat{V}_{\H xc}^{\l}[n_{\Phi^\l}]$ is the previously introduced local Hartree-exchange-correlation potential operator evaluated at the DS1H density $n_{\Phi^\l}$. Evidently, in practice, Eq.~(\ref{DS1Heigenval}) is decomposed into usual one-particle hybrid KS equations. For simplicity, we will now refer to this DS1H wave function and associated density as just $\Phi$ and $n$, respectively, the $\l$-dependence being implicit. The DS1H energy is thus finally written as
\begin{eqnarray}
E^{\text{DS1H},\l} &=& \bra{\Phi} \hat{T}+\hat{V}_{\text{ext}} \ket{\Phi} + E_{\H}[n] + \l E_x^{\HF}[\Phi]
\nonumber\\
&& + (1-\l) E_x[n] + E_c[n] -\l^2 E_c[n_{1/\l}],
\label{DS1H}
\end{eqnarray} 
where the full Coulombic Hartree energy $E_{\H}[n]$ has been recomposed, $E_x^{\HF}[\Phi]$ is the HF exchange energy and in practice density-functional approximations must be used for $E_x[n]$ and $E_c[n]$. In the appendix, we give the explicit formulas for calculating the scaled functional $E_c[n_{1/\l}]$ and associated potential for LDA and GGA approximations. Neglecting the density scaling in the correlation functional, $E_c[n_{1/\l}]\approx E_c[n]$, in Eq.~(\ref{DS1H}) gives an {\it one-parameter hybrid} (1H) approximation,
\begin{eqnarray}
E^{\text{1H},\l} &=& \bra{\Phi} \hat{T}+\hat{V}_{\text{ext}} \ket{\Phi} + E_{\H}[n] + \l E_x^{\HF}[\Phi]
\nonumber\\
&& + (1-\l) E_x[n] + (1 -\l^2) E_c[n],
\label{1H}
\end{eqnarray} 
which has a similar form than the standard one-parameter hybrid functionals such as B1LYP~\cite{AdaBar-CPL-97} or PBE1PBE (also known as PBE0)~\cite{ErnScu-JCP-99a,AdaBar-JCP-99}, except that the correlation energy in Eq.~(\ref{1H}) is weighted by $(1 -\l^2)$ while in the standard one-parameter hybrid functionals it is weighted by a factor of $1$.

All what is missing in Eq.~(\ref{DS1H}) is the correlation energy associated with the scaled interaction $\l\hat{W}_{ee}$. It can be calculated by a nonlinear Rayleigh-Schr\"odinger perturbation theory~\cite{AngGerSavTou-PRA-05,FroJen-PRA-08,Ang-PRA-08} starting from the DS1H reference. Consider the following energy expression with the perturbation parameter $\alpha$:
\begin{eqnarray}
E^{\l,\alpha} &=& \min_{\Psi}\Bigl\{\bra{\Psi}\hat{T}+\hat{V}_{\text{ext}}+\l \hat{V}_{\H x}^\HF[\Phi]  + \alpha \l \hat{W} \ket{\Psi}
\nonumber\\
 &&   +\bar{E}_{\H xc}^{\l}[n_{\Psi}]\Bigl\},
\label{Ela}
\end{eqnarray}
where $\l\hat{W}=\l \left( \hat{W}_{ee} - \hat{V}_{\H x}^\HF[\Phi] \right)$ is the scaled M{\o}ller-Plesset fluctuation perturbation operator. For $\alpha=1$, Eq.~(\ref{Ela}) reduces to Eq.~(\ref{EminPsi}), so $E^{\l,\alpha=1}$ is the exact energy, independently of $\l$. The sum of the zeroth-order energy and first-order energy correction gives simply the DS1H energy, $E^{\text{DS1H},\l}=E^{\l,(0)}+E^{\l,(1)}$. Thanks to the existence of a Brillouin theorem just like in standard M{\o}ller-Plesset perturbation theory (see Refs.~\onlinecite{AngGerSavTou-PRA-05,FroJen-PRA-08,Ang-PRA-08}), only double excitations contribute to the second-order energy correction which has a standard MP2 form,
\begin{eqnarray}
E^{\l,(2)} &=& \l^2 \sum_{i<j\atop a<b} \frac{\left|\bra{ij}\ket{ab}\right|^2}{\varepsilon_{i} +\varepsilon_{j} -\varepsilon_{a} -\varepsilon_{b}} = \l^2 E_c^{\text{MP2}},
\label{MP2}
\end{eqnarray}
where $i,j$ and $a,b$ refer to occupied and virtual DS1H spin-orbitals, respectively, with associated orbital eigenvalues $\varepsilon_{k}$, and $\bra{ij}\ket{ab}$ are the antisymmetrized two-electron integrals. Note that the dependence on $\l$ is not simply quadratic since the spin-orbitals and their eigenvalues implicitly depend on $\l$. Our final {\it density-scaled one-parameter double-hybrid} (DS1DH) approximation is then obtained by adding the second-order correction to the DS1H energy
\begin{eqnarray}
E^{\text{DS1DH},\l} = E^{\text{DS1H},\l} + E^{\l,(2)}.
\label{DS1DH}
\end{eqnarray}

To summarize, the exchange-correlation energy in the DS1DH approximation is
\begin{eqnarray}
E^{\text{DS1DH},\l}_{xc} &=& \l E_x^{\HF} + (1-\l) E_x[n] 
\nonumber\\
&&+ E_c[n] -\l^2 E_c[n_{1/\l}] + \l^2 E_c^{\text{MP2}}.
\label{ExcDS1DH}
\end{eqnarray} 
To make connection with the standard double-hybrid approximations, we also define an {\it one-parameter double-hybrid} (1DH) approximation, obtained by neglecting the density scaling in the correlation functional, $E_c[n_{1/\l}]\approx E_c[n]$,
\begin{eqnarray}
E^{\text{1DH},\l}_{xc} &=& \l E_x^{\HF} + (1-\l) E_x[n] 
\nonumber\\
&&+ (1-\l^2) E_c[n] + \l^2 E_c^{\text{MP2}},
\label{Exc1DH}
\end{eqnarray} 
which exactly corresponds to the double-hybrid approximation of Eq.~(\ref{DH}) with parameters $a_x=\l$ and $a_c=\l^2$.

\begingroup
\squeezetable
\begin{table*}[t]
\caption{MAEs and MEs (in kcal/mol) on the AE6 and BH6 test sets for several methods. For the single-hybrid DS1H, 1H, PBE1PBE, and B1LYP approximations and the double-hybrid DS1DH and 1DH approximations, the results are for the optimal values of $\l$ which minimize the MAEs of the AE6 and BH6 sets, separately. For the range-separated hybrids, this is the range-separation parameter $\mu$ (in bohr$^{-1}$) which is optimized. All calculations were carried out with the cc-pVQZ basis set.}
\label{tab:AE6BH6}
\begin{tabular}{llrrcllrr}
\hline
\hline
          & \multicolumn{3}{c}{AE6}           &                &  \multicolumn{3}{c}{BH6}  \\
          \cline{2-4} \cline{6-8}
Method    & $\l$ or $\mu$     &MAE  &  ME     & \phantom{xxxx} &  $\l$ or $\mu$& MAE  &  ME  \\
\hline                                                  
HF        &                   & 145.1& -145.1  &                &            & 12.2 & 12.2 \\
LDA       &                   & 76.9 & 76.9  &                &            & 18.0 &-18.0 \\
PBE       &                   & 15.5 & 12.4  &                &            & 9.61 &-9.61 \\
BLYP      &                   & 6.52 & -1.18  &                &            & 8.10 &-8.10 \\
MP2       &                   & 6.86 &  4.17  &                &            & 3.32 & 3.11 \\
\\
\textit{Single-hybrid approximations} & & & & & & & \\
DS1H-LDA   &  $\l=0.45$        & 5.90 & -5.20 &                &  $\l=0.50$ & 1.62 & -1.62 \\
1H-LDA   &  $\l=0.45$        & 7.04 & 1.38  &                &  $\l=0.60$ & 1.98 & -0.59 \\
DS1H-PBE & $\l=0.20$        & 5.18 &  -3.85  &                &  $\l=0.45$ & 1.00 & 0.35 \\
1H-PBE   & $\l=0.20$         & 4.43 & -2.00  &                &  $\l=0.45$ & 0.97 &-0.02 \\
DS1H-BLYP& $\l=0.05$         & 5.71 & -3.72  &                &  $\l=0.35$ & 1.70 &-0.15 \\ 
1H-BLYP & $\l=0.05$         & 5.62 & -3.53 &                &  $\l=0.40$ & 1.80 &-0.24 \\
PBE1PBE &  $\l=0.30$       & 5.28 &  -2.06  &                &  $\l=0.55$  & 1.22 &-0.01 \\
B1LYP   &  $\l=0.05$       & 5.52 &  -3.25  &                &  $\l=0.45$  & 1.94 &-0.77  \\          
B3LYP   &                  & 2.51 &  -1.95  &                &             & 4.95 &-4.95 \\
RSHX-PBE(GWS) &      $\mu=0.65$  & 7.78 & 1.08   &           & $\mu=0.55$ & 1.82 &-0.33 \\
RSHX-PBE(HSE) = LC-$\omega$PBE \phantom{xxxx} &      $\mu=0.40$  & 4.57 & -1.96   &           & $\mu=0.45$ & 1.09 &-0.25 \\
\\
\textit{Double-hybrid approximations} & & & & & & & \\
DS1DH-LDA & \multicolumn{3}{c}{no minimum$^*$}    &                &  $\l=0.85$ & 1.59 & 0.22 \\
1DH-LDA   & \multicolumn{3}{c}{no minimum$^*$}    &                &  $\l=0.90$ & 2.29 & 0.27 \\
DS1DH-PBE & $\l=0.65$         & 3.78 & 1.30  &                &  $\l=0.80$ & 1.32 & 0.48\\
1DH-PBE   & $\l=0.55^\dag$    & 8.64$^\dag$ & 7.06$^\dag$  &                &  $\l=0.80$ & 1.42 & 0.12 \\
DS1DH-BLYP & $\l=0.80$        & 4.73 &  -2.52  &                &  $\l=0.65$ & 0.60 & 0.24 \\
1DH-BLYP   & $\l=0.55$         & 1.46 & 0.07  &                &  $\l=0.75$ & 0.80 &-0.18 \\
B2-PLYP   &                   & 1.39 & -1.09   &                &            & 2.21 &-2.21 \\
RSH-PBE(GWS)+lrMP2 & $\mu=0.50$     & 3.48 & -1.91   &                &  $\mu=0.70$& 1.55 & 0.73 \\
\hline
\hline
\multicolumn{8}{l}{$^*$ There is no minimum for $0<\lambda<1$. The global minimum is for $\l=1.0$, i.e. MP2.}\\
\multicolumn{8}{l}{$^\dag$ This is a local minimum. The global minimum is for $\l=1.0$, i.e. MP2.}
\end{tabular}
\end{table*}
\endgroup

\section{Computational details}

Except for the B2-PLYP calculations which were carried out with GAUSSIAN 09~\cite{Gaussian-PROG-09}, all other calculations have been performed with a development version of the MOLPRO 2008 program~\cite{Molproshort-PROG-08}, in which the DS1DH and 1DH approximations have been implemented. For $E_x[n]$ and $E_{c}[n]$, we use the LDA functional~\cite{VosWilNus-CJP-80} and two GGA functionals, Perdew-Burke-Ernzerhof (PBE)~\cite{PerBurErn-PRL-96} and BLYP~\cite{Bec-PRA-88,LeeYanPar-PRB-88}. For DS1DH approximations, the corresponding density-scaled correlation energy is obtained from the formulas of the appendix. For a given value of the parameter $\l$, a self-consistent hybrid calculation is first performed and the MP2 correlation energy part calculated with the obtained orbitals is then added. The empirical parameter $\l$ is optimized on the AE6 and BH6 test sets~\cite{LynTru-JPCA-03}. The AE6 set is a small representative benchmark set of six atomization energies consisting of SiH$_4$, S$_2$, SiO, C$_3$H$_4$ (propyne), C$_2$H$_2$O$_2$ (glyoxal), and C$_4$H$_8$ (cyclobutane). The BH6 set is a small representative benchmark set of forward and reverse hydrogen barrier heights of three reactions, OH + CH$_4$ $\to$ CH$_3$ + H$_2$O, H + OH $\to$ O + H$_2$, and H + H$_2$S $\to$ HS + H$_2$. We compute mean errors (MEs) and mean absolute errors (MAEs) as functions of the parameter $\l$. All the calculations for the AE6 and BH6 sets were performed at the geometries optimized by quadratic configuration interaction with single and double excitations with the modified Gaussian-3 basis set (QCISD/MG3)~\cite{LynZhaTru-JJJ-XX}. The best double-hybrid approximations are also compared on larger benchmark sets which consist of the set of 49 atomization energies of Ref.~\onlinecite{FasCorSanTru-JPCA-99} (G2-1 test set~\cite{CurRagTruPop-JCP-91,CurRagRedPop-JCP-97} except for the six molecules containing Li, Be, and Na) at MP2(full)/6-31G* geometries, and the DBH24/08 test set~\cite{ZheZhaTru-JCTC-07,ZheZhaTru-JCTC-09} of 24 forward and reverse reaction barrier heights with QCISD/MG3 geometries. One practical advantage of these benchmark sets is that, as for the AE6 and BH6 sets, they come with reference values with zero-point energies removed and which can therefore be directly compared to the differences of electronic energies. We use the Dunning cc-pVTZ, cc-pVQZ, and aug-cc-pVQZ basis sets~\cite{Dun-JCP-89,KenDunHar-JCP-92,WooDun-JCP-93,WooDun-JCP-94}. Core electrons are kept frozen in all our MP2 calculations. Spin-restricted calculations are performed for all the closed-shell systems, and spin-unrestricted calculations for all the open-shell systems.

\section{Results and discussion}

\begingroup
\squeezetable
\begin{table*}[t]
\caption{MAEs and MEs (in kcal/mol) on the AE6 and BH6 test sets for the DS1DH-PBE and RSH-PBE(GWS)+lrMP2 approximations with the cc-pVTZ and cc-pVQZ basis sets. The results are for the optimal values of $\l$ or $\mu$ which minimize the MAEs of the AE6 and BH6 sets, separately.}
\label{tab:VTZVQZ}
\begin{tabular}{lllrrcllrr}
\hline
\hline
        &   & \multicolumn{3}{c}{AE6}           &                &  \multicolumn{3}{c}{BH6}  \\
          \cline{3-5} \cline{7-9}
Method             & basis \phantom{xxxxxxx} &   $\l$ or $\mu$     &MAE  &  ME     & \phantom{xxxx} &  $\l$ or $\mu$& MAE  &  ME  \\
\hline
DS1DH-PBE          & cc-pVTZ   &   $\l=0.60$        & 3.91 & -3.71  &            &   $\l=0.75$         & 1.15 & 0.02 \\
                   & cc-pVQZ   &   $\l=0.65$        & 3.78 & 1.30  &            &   $\l=0.80$         & 1.32 & 0.48 \\
RSH-PBE(GWS)+lrMP2 & cc-pVTZ   &   $\mu=0.50$       & 4.72 & -3.84  &            &   $\mu=0.70$         & 1.45 & 0.78 \\
                   & cc-pVQZ   &   $\mu=0.50$       & 3.48 & -1.91  &            &   $\mu=0.70$        & 1.55 & 0.73 \\
\hline
\hline
\end{tabular}
\end{table*}
\endgroup

Figure~\ref{fig:AE6BH6} shows the MAEs for the AE6 and BH6 test sets as functions of the parameter $\l$ for the DS1DH and 1DH approximations with the LDA, PBE, and BLYP exchange-correlation density functionals. For $\l=0$, each double-hybrid approximation reduces to the corresponding standard Kohn-Sham density-functional approximation. For $\l=1$, all our double-hybrid approximations reduce to MP2 with HF orbitals. The MAEs and MEs at the optimal values of $\l$ which minimize the MAEs on the AE6 and BH6 sets are also reported in Table~\ref{tab:AE6BH6}, and compared to those obtained with standard HF, LDA, PBE, BLYP, and MP2, as well as with other hybrid approximations.

Let us start by discussing the double-hybrid results for the AE6 atomization energies. Both DS1DH-LDA and 1DH-LDA have a larger MAE than MP2 for all $\l<1$, which makes these double-hybrid approximations of little value. While 1DH-PBE also gives a larger MAE than MP2 for all $\l<1$, DS1DH-PBE is much more accurate than both standard PBE and MP2 near the optimal parameter value of $\l=0.65$. In contrast, DS1DH-BLYP appears to be less accurate than 1DH-BLYP. On the AE6 set, the latter leads to the smallest MAE among our one-parameter double-hybrid approximations with a minimal MAE of 1.46 kcal/mol for the optimal parameter $\l=0.55$. The B2-PLYP double-hybrid gives a similar MAE of 1.39 kcal/mol (though with a ME farther away from zero). In fact, 1DH-BLYP is just a one-parameter version of the B2-PLYP double hybrid with optimal parameters $a_x=\l=0.55$ and $a_c=\l^2\simeq 0.30$ very close to the original B2-PLYP parameters $a_x=0.53$ and $a_c=0.27$.

The fact that neglecting the density scaling in the correlation functional, $E_c[n_{1/\l}]\approx E_c[n]$ (i.e., going from Eq.~(\ref{ExcDS1DH}) to Eq.~(\ref{Exc1DH})), yields a greater accuracy on the AE6 set for the double-hybrid approximation based on the BLYP functional but worsen the double-hybrid approximation based on the PBE functional can be clarified by looking at the (signed) MEs. It appears that, in both cases, the 1DH approximation always gives a more positive ME in comparison to the DS1DH approximation, at the optimal values of $\l$, and in fact also for all $\l$ (not shown). Since DS1DH-PBE gives a positive ME (1.30 kcal/mol) at the optimal $\l$, inherited from the large positive ME of standard PBE (12.4 kcal/mol), it follows that neglecting density scaling makes the ME even more positive (7.06 kcal/mol), thus deteriorating the accuracy of this double hybrid. On the contrary, since DS1DH-BLYP gives a negative ME (-2.52 kcal/mol) at the optimal $\l$, inherited from the negative ME of standard BLYP (-1.18 kcal/mol), neglecting density scaling makes the ME vary in the right direction, reaching a ME of 0.07 kcal/mol and also improving the MAE.

Let us consider now the double-hybrid results for the BH6 barrier heights. The MAE curves of all the DS1DH and 1DH approximations now display a marked minimum at an intermediate value of $\l$, thus improving upon both the corresponding standard Kohn-Sham density-functional approximations and MP2. In comparison to AE6, the minimal MAEs for the BH6 set are obtained for larger values of $\l$, from 0.65 to 0.90, which is consistent with the commonplace experience that a larger fraction of HF exchange improves barrier heights (by decreasing the self-interaction error). For this BH6 set, the DS1DH approximations are found to give smaller MAEs than the 1DH approximations for all the three density-functional approximations tested here. The best double-hybrid approximation is DS1DH-BLYP with a minimal MAE of 0.60 kcal/mol at $\l=0.65$. The B2-PLYP double-hybrid gives a larger MAE of 2.21 kcal/mol, but it has not been optimized for barrier heights.

For each of our three best one-parameter double-hybrid approximations, we have also determined a global optimal value of $\l$ which minimizes the total MAE of the combined AE6+BH6 set, and which could be used in general applications: $\l=0.65$ for DS1DH-PBE giving a total MAE of 2.77 kcal/mol, $\l=0.70$ for DS1DH-BLYP giving a total MAE of 2.94 kcal/mol, and $\l=0.65$ for 1DH-BLYP giving a total MAE of 1.75 kcal/mol. Note that the optimal fractions of HF exchange and MP2 correlation in 1DH-BLYP, $a_x=\l=0.65$ and $a_c = \l^2 \approx 0.42$, roughly reproduce the two parameters of the B2GP-PLYP double hybrid, i.e. $a_x=0.65$ and $a_c=0.36$.

\begingroup
\squeezetable
\begin{table*}[t]
\caption{Atomization energies (in kcal/mol) of the 49 molecules of the set of Ref.~\onlinecite{FasCorSanTru-JPCA-99} (G2-1 test set except for the six molecules containing Li, Be, and Na). The calculated values were obtained using the double hybrids 1DH-BLYP and B2-PLYP, and the range-separated double hybrid RSH-PBE(GWS)+lrMP2 with the cc-pVQZ basis set and MP2(full)/6-31G* geometries. The results are for the optimal value of $\l=0.65$ for 1DH-BLYP and the optimal value of $\mu=0.58$ for RSH-PBE(GWS)+lrMP2 which minimize the total MAE of the combined AE6+BH6 set. The zero-point energies are removed in the reference values. For each method, the value with the largest error is indicated in boldface.}
\label{tab:G55}
\begin{tabular}{lcccc}
\hline
\hline

Molecule                 & \multicolumn{1}{c}{1DH-BLYP} & \multicolumn{1}{c}{B2-PLYP} & RSH-PBE(GWS)+lrMP2 & \multicolumn{1}{c}{Reference$^a$} \\
\hline
CH	                 &       83.12	   &        83.70  &  78.38         &    84.00 \\
CH$_2$ ($^{3}$B$_{1}$)	 &       190.61	   &       190.57  &  190.19        &    190.07\\
CH$_2$ ($^{1}$A$_{1}$)   &       178.51     &	   178.84  &  170.26        &    181.51\\
CH$_3$	                 &       307.74     &	   307.90  &  302.91        &    307.65\\
CH$_4$	                 &       419.70	   &       419.19  &  410.84        &    420.11\\
NH	                 &        83.83	   &        84.89  &  81.09         &    83.67 \\
NH$_2$	                 &       183.02	   &       183.94  &  177.12        &   181.90\\
NH$_3$	                 &       297.74	   &       297.69  &  288.76        &   297.90 \\
OH	                 &       106.64	   &       106.43  & 104.49         &   106.60\\
OH$_2$	                 &        231.43   &       229.81  & 225.48         &   232.55\\
FH	                 &        140.47   &       139.00  & 137.20         &   141.05\\
SiH$_2$ ($^{1}$A$_{1}$)	 &       151.11	   &       151.77  &  143.21        &   151.79  \\
SiH$_2$ ($^{3}$B$_{1}$)	 &       131.56	   &       131.78  &  133.05        &  131.05\\
SiH$_3$                  &        226.10   &       226.67  &  220.05        &  227.37 \\
SiH$_4$                  &       321.41	   &       321.95  &  311.89        & 322.40  \\
PH$_2$	                 &       153.32	   &       154.80  &  146.37        & 153.20  \\
PH$_3$                   &       239.45	   &       240.73  &  229.18        & 242.55 \\
SH$_2$                   &       180.83	   &       180.58  &  174.18        &  182.74 \\
ClH	                 &        105.69   &       105.01  &  101.63        &  106.50 \\
HCCH	                 &         406.96   &      404.45   &   399.05       & 405.39  \\
H$_2$CCH$_2$	         &        563.66   &      562.15   &    554.55      & 563.47 \\
H$_3$CCH$_3$             &       711.68	   &      710.22   &   701.94       &  712.80 \\
CN	                 &       180.09	   &      179.61   &  172.93        & 180.58 \\
HCN	                 &       316.61	   &      314.12   &  305.21        & 313.20 \\
CO	                 &        261.67   &      258.28   & 254.60         & 259.31 \\
HCO	                 &        282.86   &      280.62   &  277.00        & 278.39 \\
H$_2$CO	                 &        376.06   &      373.56   &  367.85        & 373.73 \\
H$_3$COH	         &        512.96   &     510.38	   &  505.00        & 512.90 \\
N$_2$	                 &      231.77	   & 229.24	   &   218.09       &228.46 \\
H$_2$NNH$_2$             &       439.16	   & 438.77	   &   428.92       &  438.60\\
NO	                 &     156.75	   &  155.04	   &   151.17       &  155.22 \\
O$_2$	                 &     125.03       & 122.71	   &  119.73        & 119.99  \\
HOOH	                 &    268.15	   &  265.44	   &   259.77       & 268.57 \\
F$_2$	                 &   38.01	   & 36.29	   &   31.64        & 38.20  \\
CO$_2$	                 &   {\bf 396.70}  &  391.23	   &   390.46       & 389.14 \\
Si$_2$	                 &   71.33	   & 70.58	   &   67.21        & 71.99 \\
P$_2$	                 &   116.60	   &   115.84	   &  107.27        & 117.09 \\
S$_2$	                 &   103.29	   &  102.27	   &  100.90        & 101.67 \\
Cl$_2$                   &    56.89	   & 55.48	   &  54.19         & 57.97 \\
SiO	                 &   193.94	   & 190.82	   &  185.82        &  192.08 \\
SC	                 &   171.31	   &  168.86	   & 163.07         &  171.31 \\
SO	                 &   127.14	   & 125.33	   &  122.46        & 125.00 \\
ClO	                 &    62.63	   &   62.70	   &  60.81         &  64.49 \\
ClF	                 &   61.37	   &  59.85	   &  57.94         & 61.36 \\
Si$_2$H$_6$              &  528.92	   &  529.02       &  {\bf 517.07}  & 530.81 \\
CH$_3$Cl	         &  394.30	   & 392.62	   &   388.23       & 394.64 \\
CH$_3$SH	         & 472.15	   & 470.71	   & 463.90         & 473.84 \\
HOCl	                 & 164.43	   & 162.27	   & 158.46         & 164.36 \\
SO$_2$	                 & 257.24	   & {\bf 251.10}  & 244.46         &257.86 \\
\hline                                                                 
MAE                      &  1.4            &  1.6         & 6.5           &      \\         
ME                       &  0.3           &  -1.0         & -6.4           &      \\ 
\hline
\hline
$^a$From Ref.~\onlinecite{FasCorSanTru-JPCA-99}.\\
\end{tabular}
\end{table*}
\endgroup

For comparison, we have also reported in Table~\ref{tab:AE6BH6} the MAEs and MEs obtained with the single-hybrid DS1H and 1H approximations, as well as the usual single-hybrid functionals PBE1PBE~\cite{ErnScu-JCP-99a,AdaBar-JCP-99} and B1LYP~\cite{AdaBar-CPL-97}, both with reoptimization of the fraction of HF exchange $\l$, and the standard B3LYP functional~\cite{Bec-JCP-93,BarAda-CPL-94,SteDevChaFri-JPC-94} without reoptimization of the parameters. We have also considered range-separated single-hybrid functionals (also known as long-range corrected functionals~\cite{IikTsuYanHir-JCP-01}), here referred to as RSHX as in Ref.~\onlinecite{GerAng-CPL-05a}, with two short-range exchange PBE functionals, the Goll-Werner-Stoll (GWS) one~\cite{GolWerSto-PCCP-05,GolWerStoLeiGorSav-CP-06} (which is a modified version of the one of Ref.~\onlinecite{TouColSav-JCP-05}) and the Heyd-Scuseria-Ernzerhof (HSE) one~\cite{HeyScuErn-JCP-03}. For notational consistency, we refer to these two range-separated single-hybrid functionals as RSHX-PBE(GWS) and RSHX-PBE(HSE), respectively, although RSHX-PBE(HSE) is in fact known in the literature as LC-$\omega$PBE~\cite{VydScu-JCP-06}. In the case of range separation, this is the (nonlinear) inverse range parameter $\mu$ which plays the role of $\l$ and which is optimized. The single hybrids DS1H-PBE and 1H-PBE give very similar results than PBE1PBE. The same is true for DS1H-BLYP and 1H-BLYP in comparison with B1LYP. It appears that PBE1PBE is less accurate than DS1DH-PBE on the AE6 set and about as accurate on the BH6 set. The single hybrids B1LYP and B3LYP are also found to be significantly less accurate than the best one-parameter double-hybrid approximation constructed with the BLYP functional, namely 1DH-BLYP, on both the AE6 and BH6 sets. RSHX-PBE(HSE) gives slightly smaller MAEs than PBE1PBE, but it is still less accurate than DS1DH-PBE on the AE6 set and only slightly more accurate on the BH6 set. Quite unexpectedly, RSHX-PBE(GWS) is much less accurate than PBE1PBE, which points to a weakness of the short-range exchange PBE functional of Ref.~\onlinecite{GolWerStoLeiGorSav-CP-06}, at least when combined with the standard (full-range) correlation PBE functional. All these results globally confirm the greater potentiality of double hybrids over single hybrids.

\begingroup
\squeezetable
\begin{table*}[t]
\caption{Forward (F) and reverse (R) reaction barrier heights (in kcal/mol) that constitute the DBH24/08 test set. The calculated values were obtained using the double hybrids 1DH-BLYP and B2-PLYP, and the range-separated double hybrid RSH-PBE(GWS)+lrMP2 with the aug-cc-pVQZ basis set and QCISD/MG3 geometries. The results are for the optimal value of $\l=0.65$ for 1DH-BLYP and the optimal value of $\mu=0.58$ for RSH-PBE(GWS)+lrMP2 which minimize the total MAE of the combined AE6+BH6 set. The zero-point energies are removed in the reference values. For each method, the value with the largest error is indicated in boldface.}
\label{tab:Tru24}
\begin{tabular}{lcccc}
\hline
\hline

Reactions & 1DH-BLYP & B2-PLYP & RSH-PBE(GWS)+lrMP2 & Reference$^a$ \\
\hline

                                                                       &    F/R       &     F/R      &   F/R       & F/R                    
\\                                                                                                    
\textit{Heavy-atom transfer} \\                                                         
H+N$_{2}$O $ \xrightarrow ~ $ OH +N$_{2}$                              & 19.15/{\bf 77.74} & 16.53/{\bf 77.02}  & 19.34/{\bf 77.14} & 17.13/82.47 \\
H+ClH $ \xrightarrow ~ $  HCl + H                                      & 17.26/17.26  & 15.94/15.94  & 19.77/19.77 & 18.00/18.00 \\
CH$_{3}$+FCl $ \xrightarrow ~ $ CH$_{3}$F + Cl                         &  5.29/58.78  &  3.02/56.24  &  8.22/64.21 &  6.75/60.00  \\
\\
\textit{Nucleophilic substitution} \\
Cl$^{-...}$CH$_{3}$Cl $ \xrightarrow ~ $  {ClCH$_{3}$}$^{...}$Cl$^{-}$                & 11.55/11.55  & 10.76/10.76  & 15.40/15.40   & 13.41/13.41 \\
F$^{-...}$CH$_{3}$Cl $ \xrightarrow ~ $ {FCH$_{3}$}$^{...}$Cl$^{-} $                  &  2.20/27.25  &  1.59/27.01  &  4.72/31.46 & 3.44/29.42 \\ 
 OH$^{-}$+CH$_{3}$F $ \xrightarrow ~ $  HOCH$_{3}$ + F$^{-}$                          & -3.51/16.04 & -3.68/15.90   & -1.59/21.58 & -2.44/17.66 \\
\\
\textit{Unimolecular and association} \\
 H+N$_{2}$ $ \xrightarrow ~ $ HN$_{2}$                                             & 14.63/10.33   & 12.29/10.50   & 14.04/13.10  & 14.36/10.61 \\
 H+C$_{2}$H$_{4}$ $  \xrightarrow ~ $ CH$_{3}$CH$_{2}$                             & 2.96/43.04    & 1.77/42.53   & 2.701/45.76 & 1.72/41.75 \\
 HCN $\xrightarrow ~ $ HNC                                                         & 49.34/33.39   & 48.65/33.35  & 48.52/34.81 & 48.07/32.82 \\
\\
\textit{Hydrogen transfer} \\               
 OH+ CH$_{4}$ $ \xrightarrow ~ $ CH$_{3}$ + H$_{2}$O                               &  4.68/18.54   &  4.26/17.10   & 6.03/19.75  & 6.70/19.60  \\               
 H + OH $ \xrightarrow ~ $ O +H$_{2}$                                              &   9.96/10.80  &   8.06/9.72  & 13.44/10.03 & 10.70/13.10 \\ 
 H+ H$_{2}$S $ \xrightarrow ~ $ H$_{2}$ + HS                                       &   2.82/16.53  &   1.85/16.65 & 4.73/15.35  & 3.60/17.30   \\
\hline
MAE                                                                                &       1.4    &     2.0     &  2.1   \\
ME                                                                                 &      -0.8    &    -1.8     &  1.1    \\
\hline
\hline
$^a$From Ref.~\onlinecite{ZheZhaTru-JCTC-09}.\\  
\end{tabular}
\end{table*}
\endgroup

We discuss now the results obtained with the range-separated double-hybrid approach of Ref.~\onlinecite{AngGerSavTou-PRA-05}, using the short-range exchange-correlation PBE functional of Ref.~\onlinecite{GolWerStoLeiGorSav-CP-06}, referred to as RSH-PBE(GWS)+lrMP2. For the AE6 set, we obtain an optimal value of $\mu=0.50$ bohr$^{-1}$, corresponding to the value actually used in previous studies~\cite{AngGerSavTou-PRA-05,GerAng-CPL-05b,GerAng-JCP-07,GolLeiManMitWerSto-PCCP-08,ZhuTouSavAng-JCP-10,TouZhuAngSav-PRA-10}, and a minimal MAE of 3.48 kcal/mol. This value is only marginally better than the MAE of the DS1DH-PBE double hybrid, 3.78 kcal/mol. For the BH6 set, the optimal value $\mu=0.70$ corresponds to a larger range treated by HF exchange and MP2 correlation, and the minimal MAE of 1.55 kcal/mol is again very similar to the MAE of DS1DH-PBE, 1.32 kcal/mol. This suggests that standard double hybrids and range-separated double hybrids can be comparably accurate for atomization energies and barrier heights, provided that similar density-functional approximations are used. However, range-separated double hybrids have the advantage of having a much weaker basis dependence. This is shown in Table~\ref{tab:VTZVQZ} which reports the MAEs and MEs on the AE6 and BH6 sets for the DS1DH-PBE and RSH-PBE(GWS)+lrMP2 approximations using the cc-pVTZ and cc-pVQZ basis sets. The basis dependence is clearly seen on the MEs. The absolute differences in MEs between the two basis sets for DS1DH-PBE, 5.01 kcal/mol and 0.46 kcal/mol for AE6 and BH6, respectively, are far greater than those of RSH-PBE(GWS)+lrMP2, 1.93 kcal/mol and 0.05 kcal/mol for AE6 and BH6, respectively. Notice that, for AE6, because the ME of DS1DH-PBE changes sign when going from the cc-pVTZ to the cc-pVQZ basis, the variation of the corresponding MAE turns out to be fortuitously small, and so at first glance this hides the large basis dependence of DS1DH-PBE. Note also that the optimal value of the parameter $\l$ is more dependent on the basis than the range-separation parameter $\mu$. For RSH-PBE(GWS)+lrMP2, we have determined a global optimal value of $\mu=0.58$ which minimizes the total MAE of the combined AE6+BH6 set, giving a total MAE of 2.63 kcal/mol.

Finally, we compare our best one-parameter double-hybrid approximation 1DH-BLYP (with the optimal parameter $\l=0.65$), the standard double hybrid B2-PLYP and the range-separated double hybrid RSH-PBE(GWS)+lrMP2 (with the optimal parameter $\mu=0.58$) on larger benchmark sets, the 49 atomization energies of the set of Ref.~\onlinecite{FasCorSanTru-JPCA-99} (Table~\ref{tab:G55}) and the DBH24/08 set of 24 reaction barrier heights (Table~\ref{tab:Tru24}). For the set of atomization energies in Table~\ref{tab:G55}, 1DH-BLYP is somewhat more accurate than B2-PLYP, with a slightly smaller MAE (1.4 vs. 1.6 kcal/mol) and a significantly smaller ME (0.3 vs. -1.0 kcal/mol). By contrast, the range-separated double hybrid RSH-PBE(GWS)+lrMP2 gives a MAE as large as 6.5 kcal/mol. This is most likely due to the fact that the short-range exchange-correlation functional used is based on PBE. Indeed, the previous results for the AE6 set (Table~\ref{tab:AE6BH6}) show that PBE is a much less accurate functional than BLYP for atomization energies. Unfortunately, there is no short-range exchange-correlation functional based on BLYP available yet. In fact, it is a practical advantage of the double hybrids without range separation that they do not require development of new density functional approximations. The range-separated coupled-cluster calculations on an extension of the G2/97 set of atomization energies by Goll {\it et al.}~\cite{GolWerStoLeiGorSav-CP-06} show that a better accuracy can be reached by using a short-range exchange-correlation functional based on the TPSS functional~\cite{TaoPerStaScu-PRL-03}. For the reaction barrier heights of Table~\ref{tab:Tru24}, 1DH-BLYP is on average more accurate than B2-PLYP, with a smaller MAE (1.4 vs. 2.0 kcal/mol) and an even smaller ME (-0.8 vs. -1.8 kcal/mol). The range-separated double hybrid RSH-PBE(GWS)+lrMP2 performs about equally well for the barrier heights, with a MAE of 2.1 kcal/mol and a ME of 1.1 kcal/mol. Notice that we have used the aug-cc-pVQZ basis for this DBH24/08 set. Diffuse basis functions are indeed important for describing the charged species in the nucleophilic substitutions, particularly for 1DH-BLYP and B2-PLYP and to a lesser extent for RSH-PBE(GWS)+lrMP2. For comparison, the MAEs of these three methods with the cc-pVQZ basis are 2.1, 2.7, and 2.4 kcal/mol, respectively.

\section{Conclusions}

We have rigorously derived a class of double-hybrid approximations, combining HF exchange and MP2 correlation with a semilocal exchange-correlation density functional. These double-hybrid approximations contain only one empirical parameter and uses a density-scaled correlation energy functional. Neglecting density scaling leads to a one-parameter version of the standard double-hybrid approximations. Calculations on the representative test set of atomization energies AE6 show that in practice neglecting density scaling in an approximate functional can either make the double-hybrid method less accurate (case of PBE) or more accurate (case of BLYP). Neglecting density scaling always leads to a less accurate double-hybrid method on the representative test set of reaction barrier heights BH6, for all the density-functional approximations tested here. Our best one-parameter double-hybrid approximation, 1DH-BLYP, roughly reproduces the two parameters of the standard B2-PLYP or B2GP-PLYP double-hybrid approximations. This shows that these methods are not only empirically close to an optimum for general chemical applications but are also theoretically supported. More intensive tests on larger benchmark sets of atomization energies and reaction barrier heights confirm that the double hybrid 1DH-BLYP with a fraction of HF exchange of $\l=0.65$ can reach on average near-chemical accuracy for these properties.

The range-separation double hybrid RSH+lrMP2 using the short-range exchange-correlation PBE functional of Ref.~\onlinecite{GolWerStoLeiGorSav-CP-06} is competitive with the best global double hybrids for reaction barrier heights but gives larger errors for atomization energies. Nevertheless, the range-separated double hybrids have the advantage of a weaker basis dependence and a correct long-range behavior (important, e.g., for van der Waals interactions). One could try to improve the performance of the range-separated double hybrids for thermochemistry by either using better short-range exchange-correlation functionals (such as in Ref.~\onlinecite{GolErnMoeSto-JCP-09}), or combining them with global double hybrids in a similar way as done for exchange in the CAM-B3LYP approximation~\cite{YanTewHan-CPL-04}.

Beside providing a rigorous derivation of the double-hybrid approximations, the formalism used in this work also paves the way toward other rigorous formulation of double-hybrid methods, replacing the MP2 part by some other approaches. For example, using the random phase approximation would generate a hybrid method similar to the one proposed in Ref.~\onlinecite{RuzPerCso-JCTC-10}, or using a configuration-interaction or multiconfiguration self-consistent-field approach would lead to a hybrid method capable of dealing with static electron correlation, in a similar but alternative way to the range-separated approaches~\cite{LeiStoWerSav-CPL-97,PolSavLeiSto-JCP-02,FroTouJen-JCP-07}.

\section*{Acknowledgments}
K. S. thanks Prof. I. Othman (Atomic Energy Commission of Syria) for his support. J. T. and A. S. acknowledge support from Agence Nationale de la Recherche (contract Wademecom 07-BLAN-0272). We also thank J. G. \'Angy\'an for discussions.

\appendix
\section{Density-scaled correlation energy and potential}
\label{Scalrelat}

In this appendix, we give explicit expressions for the density-scaled correlation energy functional which appears in Eq.~(\ref{ExcDS1DH})
\begin{eqnarray}
E^{\l}_{c}[n] = \l^2 E_c[n_{1/\l}],
\end{eqnarray}
and its associated potential
\begin{eqnarray}
v_{c}^{\l}[n](\b{r}) &=& \frac{\delta{E}^{\l}_{c}[n]}{\delta n(\b{r})}.
\end{eqnarray}

\subsection{Density-scaled local-density approximations}

For local-density approximations (LDA), 
\begin{eqnarray}
E_{c,\LDA}[n] &=& \int e_{c}(n(\b{r})) d\b{r},
\label{}
\end{eqnarray}
where $e_{c}$ is the energy density, the density-scaled correlation energy is obtained as
\begin{eqnarray}
E_{c,\LDA}^{\l}[n] &=& \l^2 \int  e_{c}\left(n_{1/\l}(\b{r})\right) d\b{r} 
\nonumber\\
&=& \l^2 \int  e_c \left( n(\b{r}/\l)/\l^3 \right) d\b{r}
\nonumber\\
&=& \l^5 \int  e_c \left( n(\b{r})/\l^3 \right)d\b{r},
\end{eqnarray}
where the coordinate transformation $\b{r}\to\l\b{r}$ has been used. The associated potential is simply
\begin{eqnarray}
v_{c,\LDA}^{\l}[n](\b{r}) &=& \l^2 \frac{d e_c}{d n} \left( n(\b{r})/\l^3 \right).
\nonumber\\
\end{eqnarray}

\subsection{Density-scaled generalized-gradient approximations}

For generalized-gradient approximations (GGA), usually written as a function of the density and the square of density gradient norm $\left|\b{\n}_{\b{r}}n(\b{r})\right|^2$,
\begin{eqnarray}
E_{c,\GGA}[n] &=& \int e_{c} \left(n(\b{r}),\left|\b{\n}_{\b{r}}n(\b{r})\right|^2\right) d\b{r},
\end{eqnarray}
the density-scaled correlation energy is
\begin{eqnarray}
E_{c,\GGA}^{\l}[n] &=& \l^2 \int e_{c} \left(n_{1/\l}(\b{r}),\left|\b{\n}_{\b{r}}n_{1/\l}(\b{r})\right|^2\right) d\b{r},
\nonumber\\
                   &=& \l^2 \int e_{c} \left(n(\b{r}/\l)/\l^3,\left|\b{\n}_{\b{r}}n(\b{r}/\l)\right|^2/\l^6\right) d\b{r},
\nonumber\\
                   &=& \l^5 \int e_{c} \left(n(\b{r})/\l^3,\left|\b{\n}_{\b{r}}n(\b{r})\right|^2/\l^8\right) d\b{r},
\label{Scalrel17}
\end{eqnarray}
where the coordinate transformation $\b{r}\to\l\b{r}$, and consequently $\b{\n}_{\b{r}} \to \b{\n}_{\l \b{r}}= \b{\n}_{\b{r}}/\l$, has been used. The associated potential is
\begin{eqnarray}
v_{c,\GGA}^{\l}[n](\b{r}) = \l^2 \frac{\partial e_c}{\partial n} \left( n(\b{r})/\l^3, \left|\b{\n}_{\b{r}}n(\b{r})\right|^2/\l^8 \right)
\nonumber\\
 - 2\b{\n}_{\b{r}} \cdot \left[ \frac{1}{\l^3} \frac{\partial e_c}{\partial \left|\b{\n} n \right|^2} \left( n(\b{r})/\l^3, \left|\b{\n}_{\b{r}}n(\b{r})\right|^2/\l^8 \right) \b{\n}_{\b{r}}n(\b{r}) \right].
\nonumber\\
\end{eqnarray}

The same scaling relations apply for spin-dependent functionals $E_{c}[n_{\uparrow},n_{\downarrow}]$, i.e. the scaling of the spin densities $n_{\uparrow}$ and $n_{\downarrow}$ is the same as the scaling of the total density $n$, and the scaling of the spin-density gradients $\left|\b{\n} n_{\uparrow}\right|^2$, $\left|\b{\n} n_{\downarrow}\right|^2$, and $\b{\n} n_{\uparrow} \cdot \b{\n} n_{\downarrow}$ is the same as the scaling of total density gradient $\left|\b{\n} n\right|^2$.

\bibliographystyle{apsrev4-1}

\end{document}